\documentclass[showkeys]{revtex4}

\usepackage{epsfig}

\begin{document}

\title{High quality c-axis oriented $ZnO$ thin-film obtained 
at very low pre-heating temperature}

\author{A. M. P. Santos, and Edval J. P. Santos}

 \affiliation{Laboratory for Devices and
Nanostructures at the Departamento de Eletr\^onica e Sistemas,
Universidade Federal de Pernambuco,
Caixa Postal 7800, 50670-000, Recife-PE, Brazil\\
E-mail: edval@ee.ufpe.br .
}
 
 
 
%
 
\begin{abstract} 

Highly oriented $ZnO$ thin-film has been obtained at a very 
low pre-heating temperature, with the spin-coating sol-gel technique.
The dependence of the c-axis orientation  on the pre-heating 
temperature has been studied with experimental design and 
response surface techniques to optimize the deposition 
process with respect to c-axis orientation, and surface 
uniformity.  The optimization variables selected for this
study are: pre-heating temperature, spin-coating speed and 
number of coating layers.  The films are probed with X-ray 
diffraction and electron microscopy.

\end{abstract}  
 
%
%


\maketitle

\section{Introduction}

Zinc oxide thin films exhibit a variety of properties such as: 
semiconducting (II-VI), photoconducting, piezoelectricity, birefrigence, 
acousto-optical, transparency in the infrared region, and opto-electrical 
properties. This makes this material very interesting for theoretical and 
experimental studies~\cite{W2004,PNIH2005}. $ZnO$ films can be produced in 
various phases, such as: wurtzite (hcp), zincblende (fcc), rocksalt (fcc), 
cesium chloride (sc)~\cite{FPM2006}.  Under ambient conditions $ZnO$ 
crystallizes in the wurtzite structure, a tetrahedrally coordinated 
structure with hexagonal lattice. $ZnO$ films can crystallize in different 
orientations as a function of the deposition technology, annealing 
temperature, substrate, etc.  The most common orientation is $(002)$ 
(hexagonal wurtzite), which  presents the densest atomic packing and 
minimum surface 
energy~\cite{ZLZYZ2006,AOMGB2004,BGK1998,LWLYXLC2004,CGC2004,PBS2003}.  
Quality of the film is typically determined with regard to transparency, 
conductivity, crystalline orientation,  and surface uniformity. 
Crystalline orientation is key to achieve $ZnO$ thin-film with 
piezoelectric properties.

Due to its low cost, and capability to coat large surface 
areas~\cite{WWCCW2006}, sol-gel is the technique selected 
for this study.  One possible difficulty with sol-gel is the purity of 
the film, which can affect the electrical and optical properties of the 
film, due to the high density of carrier traps and potential barriers at 
grain boundaries~\cite{LWLYXLC2004}. However, several studies have shown 
that the optical and electrical properties could be considerably improved 
by optimized deposition 
conditions~\cite{LLZSLZKF2002,YLLLLSZF2004,YAK2005,LLSLZKF2002,RI2005}. 

Many researchers have prepared zinc oxide thin films with the
sol-gel technique. Ohyama et al.~\cite{OKY1997}  studied 
the crystallization of dip-coated sol-gel deposited films to produce 
piezoelectric $ZnO$ films for SAW applications.  They varied the pre-heating 
temperature from $200 ^oC$ up to $500 ^oC$, and annealing temperature 
from $500 ^oC$ up to $800 ^oC$. They have observed that the pre-heating
temperature of dip-coated produced films has a strong effect on the 
crystal orientation. Their best result is achieved with a pre-heating
temperature of $300 ^oC$. Bao et al.~\cite{BGK1998} also used sol-gel 
deposited films and studied the orientation by varying the annealing 
temperature from $400 ^oC$ up to $600 ^oC$, which is too high temperature
for our application. In the work of Castanedo-P\'{e}rez et 
al~\cite{PSSMGDA1999}, $Zn(CH_3COO)_2$ is used as precursor, they have
studied the effect of the annealing temperature from $200 ^oC$ to 
$450 ^oC$ (pre-heating at $100 ^oC$) on the formation of $ZnO$.  
The film obtained is not highly oriented, as  they were concerned 
with the optical transmission properties of the film.
Jiwei et al.~\cite{JLX2000} deposited
$ZnO$ films on $SiO_2/Si$ also with sol-gel with pre-heating
at $200 ^oC$, annealing from $300 ^oC$ up to $600 ^oC$, the
resulting films displayed good orientation. 
Paul et al~\cite{PBS2003} used a two-step heat treatment,
$50 ^oC$ in air, followed by $550 ^oC$ in furnace, repeating the 
cycle many times.  The annealing
was carried out in high vacuum at $400 ^oC$, with good results.
Alam et al~\cite{AC2001} used dipping solution, pre-heating 
$260 ^oC$ and annealing
from  $300 ^oC$ up to $700 ^oC$, but did not achieve a good
orientation of the films. Znaidi~\cite{ZIBSK2003} studied 
the dependence of the crystallization on the relative
concentration of zinc actetate dihydrate to monoethanolamine.
The films were pre-heated at $300^oC$ and after finishing
the coating cycles, they were annealed at $550 ^oC$
to get $(002)$ oriented films. Chakrabarti et al~\cite{CGC2004} 
varied annealing temperature from $500 ^oC$ up to $700 ^oC$.  
But their films have not shown good orientation on the various 
substrates used.  Aslan et al.~\cite{AOMGB2004} achieved good quality
films by varying the annealing temperature 
from $450 ^oC$ up to $550 ^oC$. Li et al~\cite{LWLYXLC2004} 
varied the pre-heating temperature from  $100 ^oC$ up to $500 ^oC$,
but their results were reasonable above $200 ^oC$, annealed at $600 ^oC$.
Zhang et al~\cite{ZLZYZ2006}
has achieved very high quality films by starting the deposition
process with a seed layer deposited by PLD. In their work the
pre-heating temperature varied from  $300 ^oC$ up to $600 ^oC$,
and annealed at $600 ^oC$.  Just recently, came to
our attention the work of Wang et al.~\cite{WWCCW2006}, which
also uses the sol-gel method, with pre-heating temperature from 
$300 ^oC$ up to $450 ^oC$, and annealing temperature 
from $550 ^oC$ up to $800 ^oC$. Almost the same range as
in the work of Ohyama et al.~\cite{OKY1997}, but did not
get good results.

Most authors have focused their attention on the effect of the
annealing temperature on the crystallization. And have considered
high temperatures for the pre-heating cycle.

Our objective is to apply experimental design and response surface 
techniques to produce highly oriented films, with the lowest thermal 
bugget possible. The process should be compatible with 
silicon/silicon-dioxide substrates. Such films can be useful to 
make piezoelectric coatings for integrated smart sensors.
The parameters selected for study are: pre-heating temperature, spin speed 
and number of coating layers. The paper is divided into four sections, this
introduction being the first, next the experimental procedure is
described. In the third section, the results and analysis, and 
finally the conclusions.

\section{Experimental Procedure}

\subsection{Materials preparation}

Our films were prepared by dissolving zinc acetate dyhidrate, $Zn(CH_3COO)_2$ 
($99.5\%$, Merck), in methanol ($99.9\%$, Carlo Erba) under stirring at 
$60\, ^oC$,  until a transparent and 
homogeneous solution is obtained. The microscope glass substrates are cleaned 
with neutral cleaning agent in ultrasonic bath for $20\, min$, washed in 
deionized water, acetone ($5\, min$), deionized water on ultrasonic bath 
($5\, min$), isopropanol ($5\, min$) and dry nitrogen and preserved in 
desiccator. The solution is then spun-on onto the substrate. After coating 
the required number of layers,  the $ZnO$ thin film is annealed at 
$350\, ^oC$ for $5\, hours$.  This procedure was performed on glass
substrates and on silicon wafers with a silicon dioxide layer.

\subsection{Optimization of ZnO thin film}

In order to optimize the preparation of c-axis oriented ZnO thin films, 
experimental design and response surface method are used~\cite{RI2005}. 
For this study,  two experimental designs have been carried out. 
The independent variables are: pre-heating temperature, 
spin-coating speed, and number of coating layers. Relative peak 
intensity ($I_{(002)}/(I_{(002)}+I_{(001)}+I_{(101)})$~\cite{AC2001}) is 
the dependent variable.

The first experimental design is a linear model $2^3$ factorial central 
composite design, with three central points. This leads to eleven experiments, 
which includes $8$ factorial points and $3$ central points. 

After analysis of the variable effects in the first experimental design, 
another experimental design was prepared to further optimize the process. 
The second experiment is a quadratic model $2^3$-full-factorial central 
composite design, with three central points. This leads to  $17$ experiments, 
including $6$ star points factorial. 
Statistical significance of the regression coefficients was determined 
with the $F$-test analysis of variance (ANOVA), which revealed that the 
regression is statistically significant at $95\%$ of 
confidence level ($p<0.05$).

\begin{figure}[htb]
\includegraphics[width=16pc]{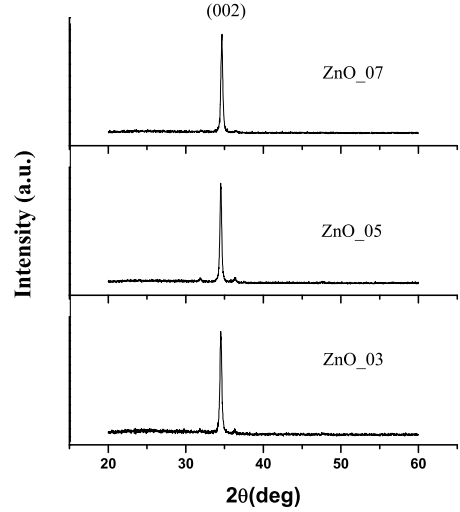}
\caption{The XRD patterns of $ZnO$ thin films annealed at $350\, ^oC$.  
$ZnO\_03$  $T_{pre}= 120\, ^oC$, $5190\, rpm$, $12\, layers$, 
$ZnO\_05$  $T_{pre}= 120\, ^oC$, $2810\, rpm$, $18\, layers$, 
$ZnO\_07$  $T_{pre}= 120\, ^oC$, $5190\, rpm$, $18\, layers$.}
\label{figura1}
\end{figure}	

\section{Results and Discussion}

The results of Ohyama et al.~\cite{OKY1997} suggest that the pre-heating 
temperature of dip-coating produced films have a strong effect on the 
crystal orientation. Recently, we have found out that, this work was 
also performed on spin-coating produced films by Wang et 
al.~\cite{WWCCW2006}. Both studies used the same temperature range.

\begin{figure}[h!tb]
\centering
\includegraphics[width=15pc]{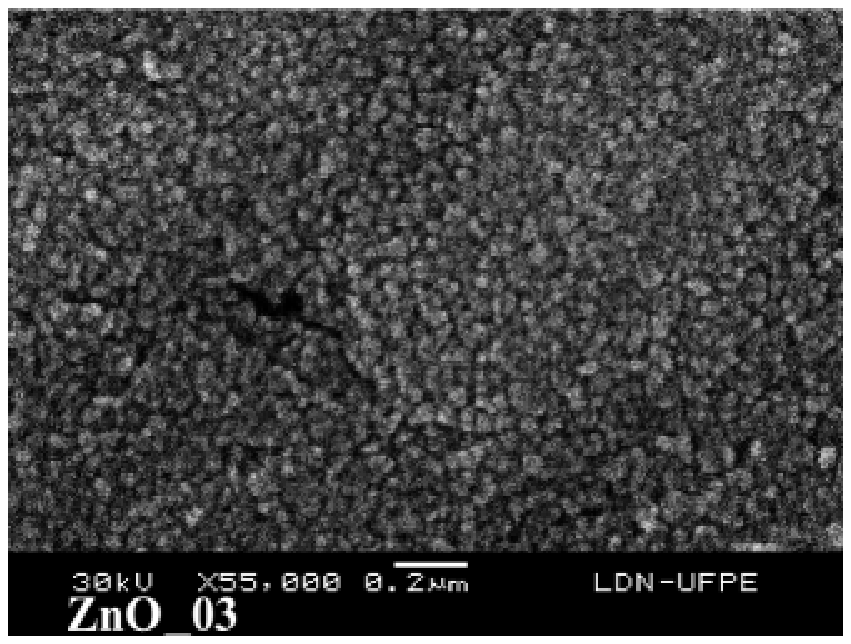}
\includegraphics[width=15pc]{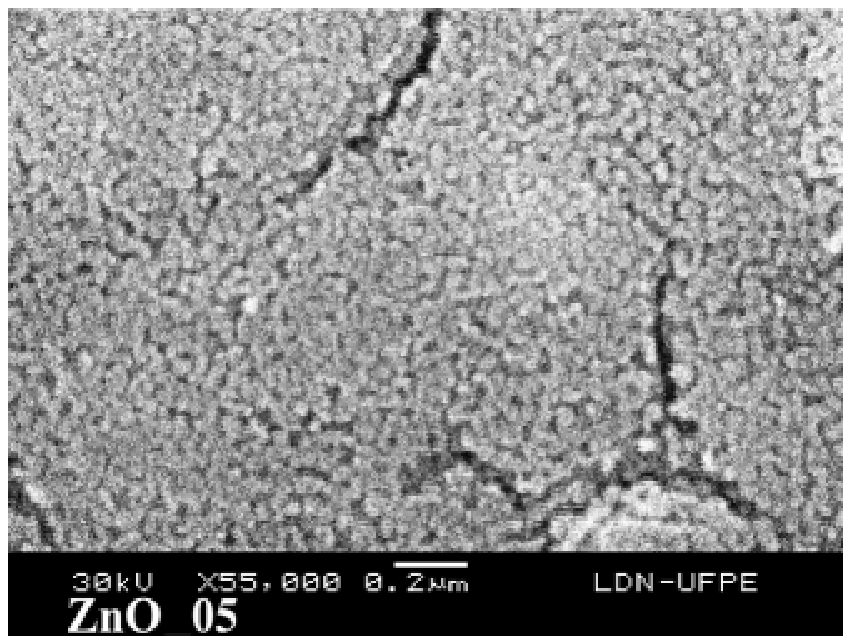}
\includegraphics[width=15pc]{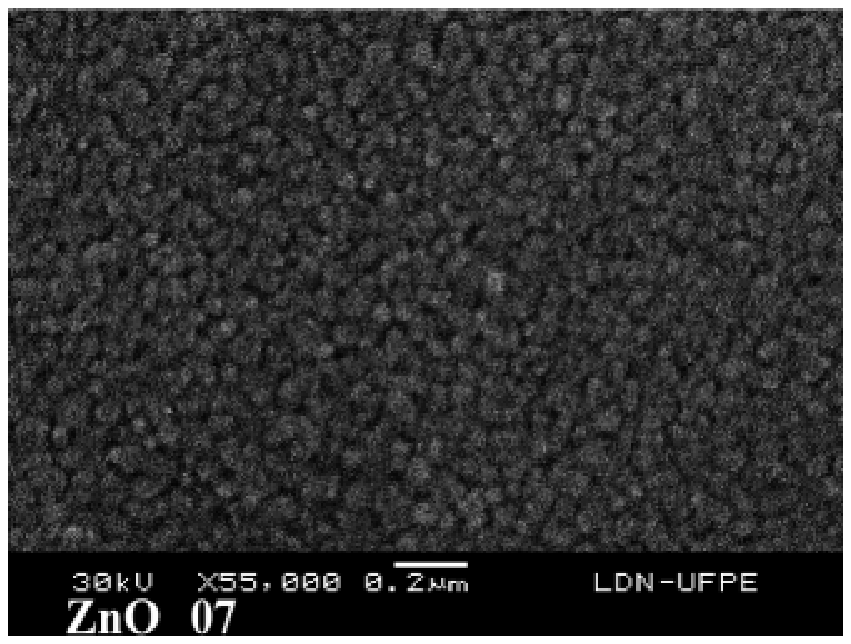}
\caption{Film surface observed with the SEM at 55,000 
magnification (bar $0.2\, \mu m$), for conditions
$ZnO\_03$  $T_{pre}= 120\, ^oC$, $5190\, rpm$, $12\, layers$, 
$ZnO\_05$  $T_{pre}= 120\, ^oC$, $2810\, rpm$, $18\, layers$, 
$ZnO\_07$  $T_{pre}= 120\, ^oC$, $5190\, rpm$, $18\, layers$.}
\label{figura2}
\end{figure}

Considering that the rate of solvent desorption and drying could
have an importante effect on the crystallization, we have decided 
to investigate this effect by using a lower temperature range for
pre-heating. Besides, we are interested in depositing very high 
quality $ZnO$ films onto silicon chips to make SAW and multilayer
BAW devices and integrated sensors, this demands the lowest temperature 
treatment possible.

\begin{figure}[htb]
\vspace{9pt}
\includegraphics[width=19pc]{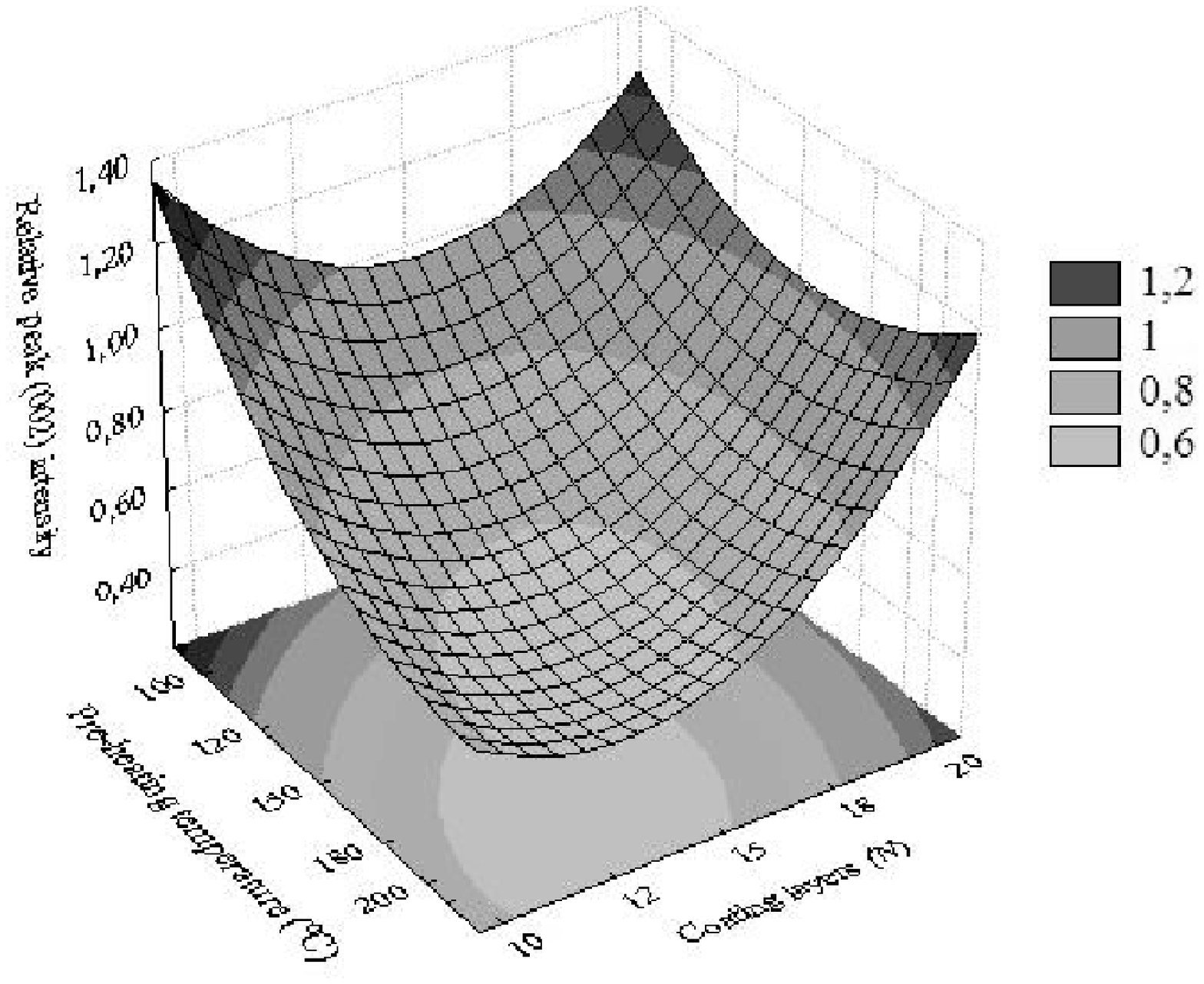}
\caption{Response surface (a) and contour diagrams (b) as 
function of pre-heating temperature and spin speed (rpm).}
\label{figura3}
\end{figure}

Following the preparation sequence presented earlier, the resulting
films were primarily observed by X-rays diffraction, with $CuKa$ 
($\lambda= 0.15418 nm$) radiation source.  The measurements 
were carried out in the range $20^o-50^o$.  The best films 
were further observed with SEM,  JEOL-6460, to exam the surface 
uniformity and grain size. 
Considering the effect of the three selected variables (pre-heating 
temperature, spin-coating speed, and number of coating layers) on the
crystallinization of the film. After the first experimental design, 
it is already clear that the pre-heating temperature is the only 
statistically significant, $(p\le0.05)$, effect.  This is also
confirmed in the second experimental design.

In Figure~1, the X-rays diffraction patterns of $ZnO$ thin films are 
presented. For the X-rays the films are deposited on glass substrates,
and annealed at $350 ^oC$ for $5\, hours$, in hot plate.
Considering the relative peak intensity, the best results for the c-axis 
orientation are obtained for
$ZnO\_03$ (pre-heating= $120\, ^oC$, spin speed= $5190\, rpm$ and coating layers= $12$),
$ZnO\_05$ (pre-heating= $120\, ^oC$, spin speed= $2810\, rpm$ and coating layers= $18$), and 
$ZnO\_07$ (pre-heating= $120\, ^oC$, spin speed= $5190\, rpm$ and coating layers= $18$).
The largest intensity of the (002)-peak is observed  at conditions 
$ZnO\_05$, while the largest relative intensity has occurred at conditions 
$ZnO\_07$. Under the SEM, $ZnO\_05$ film has displayed cracks.

\begin{table*}[htb]
\caption{(002)-peak intensity, relative peak intensity, and FWHM for the 
best results.}
\label{table:6}
\newcommand{\m}{\hphantom{$-$}}
\newcommand{\cc}[1]{\multicolumn{1}{c}{#1}}
\renewcommand{\tabcolsep}{1pc} 
\renewcommand{\arraystretch}{1.2} 
\begin{tabular}{@{}cccccc}
\hline
run     & Relative peak  & FWHM   & Grain size \\
assay   & intensity (002) & (degree) & ($nm$ (std. dev.)) \\
\hline
$ZnO\_03$ & $0.862$ & $0.25$  & $46.31$ ($7$)\\
$ZnO\_05$ & $0.877$ & $0.26$  & $58.03$ ($8$)\\
$ZnO\_07$ & $0.938$ & $0.28$  & $63.07$ ($10$)\\
\hline
\end{tabular}\\[2pt]
\end{table*}

The films with the optimal conditions also displayed the best
uniformity, as shown in  Figure~2. The grain size is of the order
of $50\,nm$, which are compatible with the FWHM measurements, as can
be seen in Table~1. The FWHM is approximately $0.26 ^o$ for all
selected samples. Considering that for the X-rays used, 
$\lambda= 0.15418\, nm $, and using Scherrer's relation 
$d= 0.94\lambda/(FWHM_{rad}\cos\theta$), the expected grain size is 
about $40\, nm$, which is close to the observed value.

Considering the relative peak intensity, a response surface plot 
was prepared.  The plot as a function of  pre-heating temperature 
and number of coating layers is presented in Figure~3. One can
see that for a low pre-heating temperature, it is possible to
get high quality film almos independent of  film thickness.

\section{Conclusions}

The methodology of experimental design and response surface analysis is
used to find the optimum process parameters for the preparation of the 
c-axis oriented $ZnO$ thin films with sol-gel. In particular, it has 
been shown that the pre-heating temperature plays an important role in 
the preparation $ZnO$ thin films with high c-axis orientation. 
Considering the range used the optimal condition parameters are:
pre-heating= $120\, ^oC$, spin speed= $5190\, rpm$ and coating layers= $18$.
With such conditions the highest $(002)$ relative peak intensity 
is achieved. We have successfully deposited c-axis oriented 
zinc oxide thin films on glass substrate from zinc acetate by 
inexpensive sol-gel process using low annealing temperature ($350 ^oC$). 
The process was also repeated on silicon dioxide.
This result is very interesting for the integration
of $ZnO$ thin films for the construction of integrated smart sensors.

\section*{Acknowledgments}

The authors thank Prof. J. A. O. de Aguiar for the X-rays measurements.
We also acknowledge the financial support of FINEP and 
PETROBRAS. One of the authors (A.M.P.S.) also acknowledges the
support of CNPq/FACEPE.

\end{document}